\documentclass[
amssymb,aps, twocolumn,prl,
showpacs,preprintnumbers,nofootinbib,
amsmath, superscriptaddress]{revtex4-2}

\usepackage{float}
\usepackage{graphicx}
\usepackage{dcolumn}
\usepackage[colorlinks=true]{hyperref}
\usepackage{color}
\usepackage{array}
\usepackage{multirow}
\usepackage[normalem]{ulem}
\usepackage{enumitem,kantlipsum}

\emergencystretch=\maxdimen
\usepackage[thinc]{esdiff}

\newcommand{\vect}[1]{\boldsymbol{#1}}

\newcommand{\comment}[1]{}

\newcommand\prlsec[1]{\vspace{2mm}\noindent \emph{#1}.---}

\emergencystretch=\maxdimen
\begin{document}

\title{Thermodynamics to infer the astrophysics of binary black hole mergers}

\author{Patrick Hu}

\affiliation{Department Physics and Astronomy, Vanderbilt University, 2301 Vanderbilt Place, Nashville, TN, 37235, USA}

\author{Karan Jani}
\affiliation{Department Physics and Astronomy, Vanderbilt University, 2301 Vanderbilt Place, Nashville, TN, 37235, USA}
\email{karan.jani@vanderbilt.edu}

\author{Kelly Holley-Bockelmann}
\affiliation{Department Physics and Astronomy, Vanderbilt University, 2301 Vanderbilt Place, Nashville, TN, 37235, USA}

\author{Gregorio Carullo}
\affiliation{Dipartimento di Fisica ``Enrico Fermi'', Universit\`a di Pisa, Pisa I-56127, Italy}
\affiliation{INFN sezione di Pisa, Pisa I-56127, Italy}


\begin{abstract}
We introduce the Merger Entropy Index ($\mathcal{I}_\mathrm{BBH}$), a new parameter to measure the efficiency of entropy transfer for any generic binary black hole merger in General Relativity. We find that $\mathcal{I}_\mathrm{BBH}$ is bounded between an asymptotic maximum and minimum. For the observed population of mergers detected by LIGO~\cite{AdvLIGORef} and Virgo~\cite{AdvVirgoRef}, we find that $\mathcal{I}_\mathrm{BBH}$ is $\lesssim30\%$ of its theoretical maximum. By imposing the thermodynamical consistency between the pre- and post-merger states through $\mathcal{I}_\mathrm{BBH}$, we showcase \texttt{BRAHMA} -- a novel framework to infer the properties and astrophysical implications of gravitational-wave detections. 
For GW190521 -- the heaviest confirmed binary black hole merger observed so far -- our framework rules out high mass-ratio, negative effective inspiral spin, and electromagnetic counterpart claims. Furthermore, our analysis provides an independent confirmation that GW190521 belongs to a separate population.
\end{abstract}

\maketitle

\prlsec{Entropy transfer} A binary black hole (BBH) merger~\cite{GW150914bbh}, regardless of its astrophysical origins, is an irreversible process. From the Second Law of Thermodynamics, the merger inevitably leads to an increase of entropy, $\Delta{S}_\mathrm{BBH}$, in the Universe:
\begin{equation}
    \Delta{S}_\mathrm{BBH} \equiv S_\mathrm{f} - ( S_1 + S_2 ) \geq 0,
\label{secondlaw}    
\end{equation}
\noindent where $S_\mathrm{f}$ is the entropy of the remnant black hole formed after merger, and $S_1$, $S_2$ are the entropies of the two pre-merger black holes.
According to the laws of black hole mechanics~\cite{1973CMaPh..31..161B}, this change in entropy is a function of the mass and spin magnitudes of black holes before $(m_{1,2}, |\vect{\chi_{1,2}}|)$ and after merger $(m_\mathrm{f}, |\vect{\chi_\mathrm{f}}|)$. For the family of Kerr black holes~\cite{PhysRevD.7.2333}, the entropy can be measured as:
\begin{equation}
    S_i = \frac{2\pi G}{c\hbar} m_i^2 \left (1 + \sqrt{1 - |\vect{\chi_i}|^2} \right) \, .
\label{entropy}    
\end{equation}

We construct a mass-invariant ratio, $\epsilon_S$, to quantify the entropy transfer during the coalescence of BBH from its pre- to post-merger stage, and note that it must be bracketed as follows:
\begin{equation}
    \epsilon_{S,\mathrm{min}} \leq \epsilon_S \equiv \frac{{S}_\mathrm{f}}{{S}_\mathrm{1} + {S}_\mathrm{2}} \leq \epsilon_{S,\mathrm{max}} \, ,
\label{kerr}    
\end{equation}
where $\epsilon_{S,\mathrm{max}}$ and $\epsilon_{S,\mathrm{min}} = 1 $ are the global maximum and minimum of entropy transfer across the BBH parameter space {$\Lambda(q,\vect{ \chi_1},\vect{\chi_2})$ with $q=m_2/m_1\leq1$} .

\prlsec{Global Maximum for BBH Entropy} By inserting Eq.~(\ref{entropy}) in Eq.~(\ref{kerr}), we find that entropy transfer has a leading order term equal to $(1+q)^2/(1+q^2)$. Therefore, the entropy transfer will be larger for equal-mass BBH mergers $(q{=}1)$.
Instead, it is in the extreme mass-ratio limit $(q \to 0)$ of BBH mergers that $\epsilon_{S} \to 1$.

To acquire an understanding of the dependence of $\epsilon_S$ on $\Lambda(q,\vect{ \chi_1},\vect{\chi_2})$, and obtain an estimate of $\epsilon_{S,\mathrm{max}}$, it is useful to first specialise to 
mergers of comparable masses $(q{\simeq}1)$ and spins $(|\vect{\chi_{1,2}}|\simeq \chi)$. In this case, the entropy transfer can be approximated as:
\begin{equation}
    \epsilon_S \approx 2\left( 1-  E_\mathrm{rad} \right)^2 \left( \frac{1+ \sqrt{1- |\vect{J} - \vect{J_\mathrm{rad}}|^2}}{1+\sqrt{1-\chi^2}} \right)  \, ,
\label{fsmax}
\end{equation} 
where $\vect{J} = \vect{\chi_1} + \vect{\chi_2} +  \vect{L}$ is the total angular momentum in the pre-merger system, $\vect{L}$ is the corresponding orbital angular momentum, and $E_\mathrm{rad}$ and $\vect{J_\mathrm{rad}}$ are the energy and angular momentum radiated to infinity via gravitational waves. The radiated quantities and angular momentum are expressed here in dimensionless units, and normalized with binary mass $(M=m_1+m_2)$.

Based on Eq.~(\ref{fsmax}), if a BBH coalescence radiates more energy, it would lead to a smaller change in entropy. If there is no residual angular momentum after the merger, it would lead to a higher change in entropy.
First noted by Hawking~\cite{PhysRevLett.26.1344}, the theoretical maximum energy that can be radiated in the comparable mass and spin case is $E_\mathrm{rad} = 1/2$. 
A minimum entropy transfer would require a binary with two maximally spinning black holes $(\chi\to 1)$ to radiate all the angular momentum $(\vect{{J}_\mathrm{rad}}~{\sim}~\vect{J})$. In this case, Eq.~(\ref{fsmax}) approaches $\epsilon_{S,\mathrm{min}}$. 
However, there exist no configurations in the $\Lambda$ hyperspace satisfying such a conditions. In fact, numerical relativity simulations find that BBH systems are unable to radiate more than ${\sim10\%}$ rest mass~\cite{2013PhRvL.111x1104M, JANI_2016, RITCatalog}. Thus, for near equal-mass BBH systems, the entropy transfer is {\it always} greater than the lower-bound imposed by the Second Law.

For generic quasi-circular BBH systems, $E_\mathrm{rad}$ and $\vect{J_\mathrm{rad}}$ are strongly dependent on the mass-ratio $q$, spin vectors  $\vect{\chi_{1,2}}$ and their orientation with respect to $\vect{L}$. The radiated quantities dictates the mass and spin of the black hole formed after the merger~\cite{Barausse:2009uz}, and consequently, the final entropy $S_\mathrm{f}$. To find the global maximum for quasi-circular BBHs, we numerically solve Eq.~(\ref{kerr}) across the $\Lambda$ hyperspace. We utilize the phenomenological fits of $E_\mathrm{rad}$ and $\vect{J_\mathrm{rad}}$~\cite{Healy:2014yta, NathanDCC, NRSur7dq4Remnant}. These fits are calibrated on numerical relativity simulations of BBH mergers~\cite{2013PhRvL.111x1104M, RITCatalog}, which incorporate the apparent horizon framework~\cite{AHF}. We find good agreement on $\epsilon_{S, \mathrm{max}}$ between the two different fits employed, $\epsilon_{S,\mathrm{max}} = 3.6235$ from~\cite{Healy:2014yta}, and $\epsilon_{S,\mathrm{max}} = 3.6342$ from~\cite{NRSur7dq4Remnant}. As expected, $\epsilon_{S,\mathrm{max}}$ occurs for a system of equal-mass, maximally spinning black holes, aligned in the direction opposite that of the orbital angular momentum.   

For the special case of head-on collisions $(L\sim0)$ of equal-mass BBHs, the entropy transfer can, in principle, be even higher. Suppose a collision occurs between two maximally spinning black holes that are anti-aligned with each other $(J\sim0)$. Provided the collision {\it radiates no energy} $(E_\mathrm{rad}~{\sim}~0)$, only then, Eq.~(\ref{fsmax}) would approach a theoretical upper-bound of $\epsilon_{S,\mathrm{max}}=4$. However, based on numerical relativity simulations, we can infer that such collisions would emit ${\gtrsim}1.35\%$ of rest mass. Thus, a subset of highly eccentric, maximally spinning, equal-mass mergers can approach an entropy transfer in the range $3.62\lesssim \epsilon_{S, \mathrm{max}} \lesssim  3.91$.

\begin{figure}[t!]
\includegraphics[scale=0.33,trim = {30 30 0 30}]{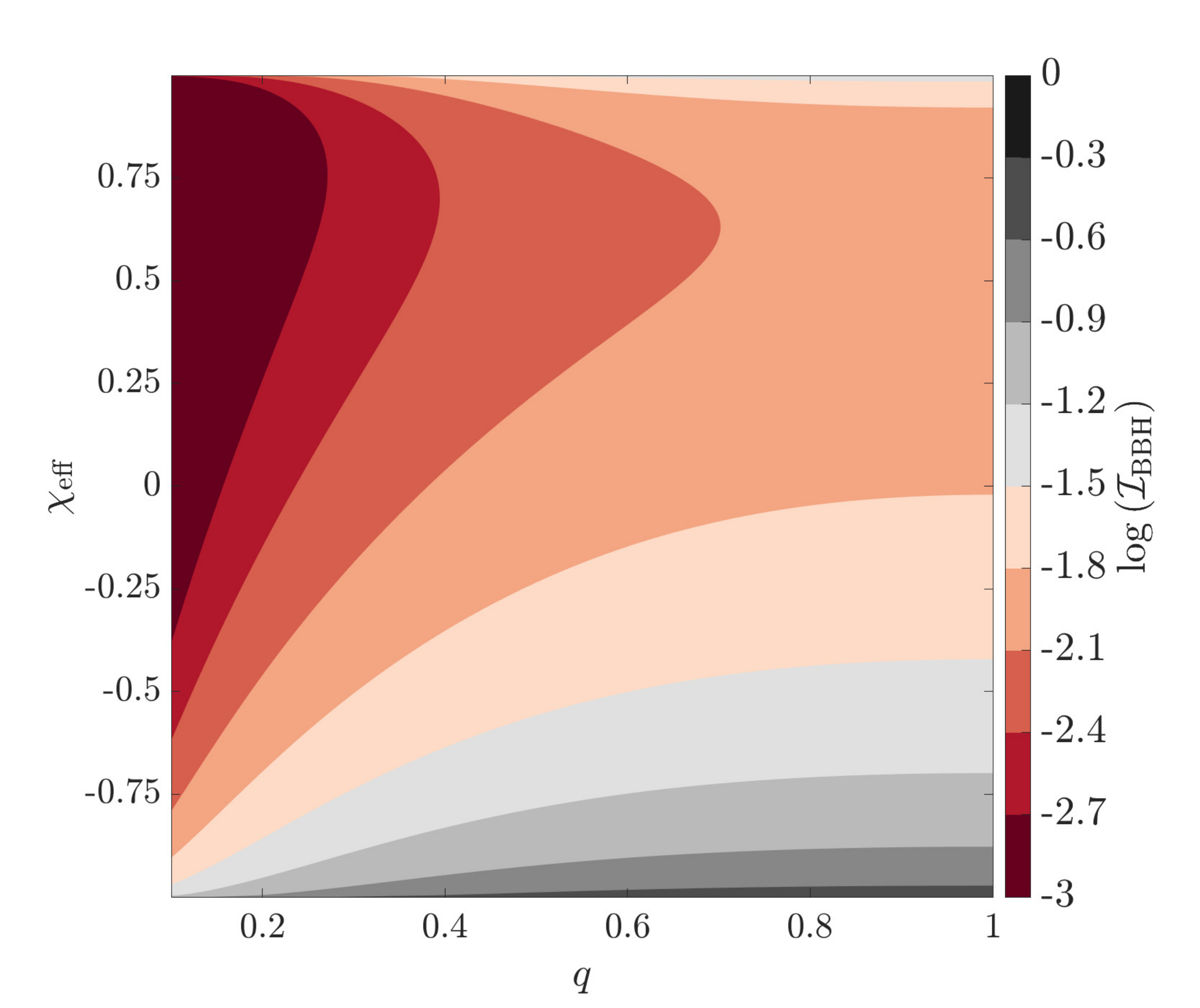}
\caption{Global behavior of Merger Entropy Index $(\mathcal{I}_\mathrm{BBH})$ {within} the intrinsic BBH parameter space, $\Lambda$. For demonstration, this figure only shows non-precessing binaries with equal spin vectors.  }
\label{fig1} 
\end{figure}

\prlsec{Merger Entropy Index}
By using the bounds of $\epsilon_S$, we construct the Merger Entropy Index as,
\begin{equation}
 \boxed{  \mathcal{I}_\mathrm{BBH}  \equiv \frac{\pi}{9} \frac{\Delta{S}_\mathrm{BBH}}{{S}_\mathrm{1} + {S}_\mathrm{2}}}  \,\, .
\label{mei}   
\end{equation}
This index measures the efficiency of entropy transfer during a BBH coalescence.  All BBH systems in General Relativity will satisfy $\mathcal{I}_\mathrm{BBH} \in (0,1]$. By construction, the index is mass-invariant and is thus applicable to a wide range of astrophysical scenarios. {For all astrophysical binaries, we anticipate  $\mathcal{I}_\mathrm{BBH}\lesssim0.9$. }

Fig.~\ref{fig1} shows the global behavior of Merger Entropy Index {within a slice in the hyperspace $\Lambda$, the intrinsic BBH parameters.} We employ \texttt{RIT} fits~\cite{Healy:2014yta} to compute $\mathcal{I}_\mathrm{BBH}$, which are valid for quasi-circular to moderately eccentric BBHs. If these BBHs are non-spinning, we find that $\mathcal{I}_\mathrm{BBH} \propto q/(1+q)^2 \equiv \eta$, the symmetric mass-ratio. 

{For near equal-mass BBHs}, the index has a rather strong dependence on the orientation of spin vectors. When the effective inspiral spin parameter $(\chi_\mathrm{eff})$~\cite{PhysRevD.84.124052} {is negative, $\mathcal{I}_\mathrm{BBH}$ exponentially approaches to 0.9.} On the contrary, when $\chi_\mathrm{eff}$ is positive, $\mathcal{I}_\mathrm{BBH}$ {has very limited variance of ${\sim}20\%$}. This is consistent regardless of the mass-ratio of the system. 
{For high mass-ratio BBHs with high primary BH spin, the index exponentially scales to the lower-bound of 0}. 
{We note that} BBHs with high in-plane spins is that they tend to have high $\mathcal{I}_\mathrm{BBH}$.

\begin{figure}[t!]
\includegraphics[scale=0.3,trim = {-20 20 0 30}]{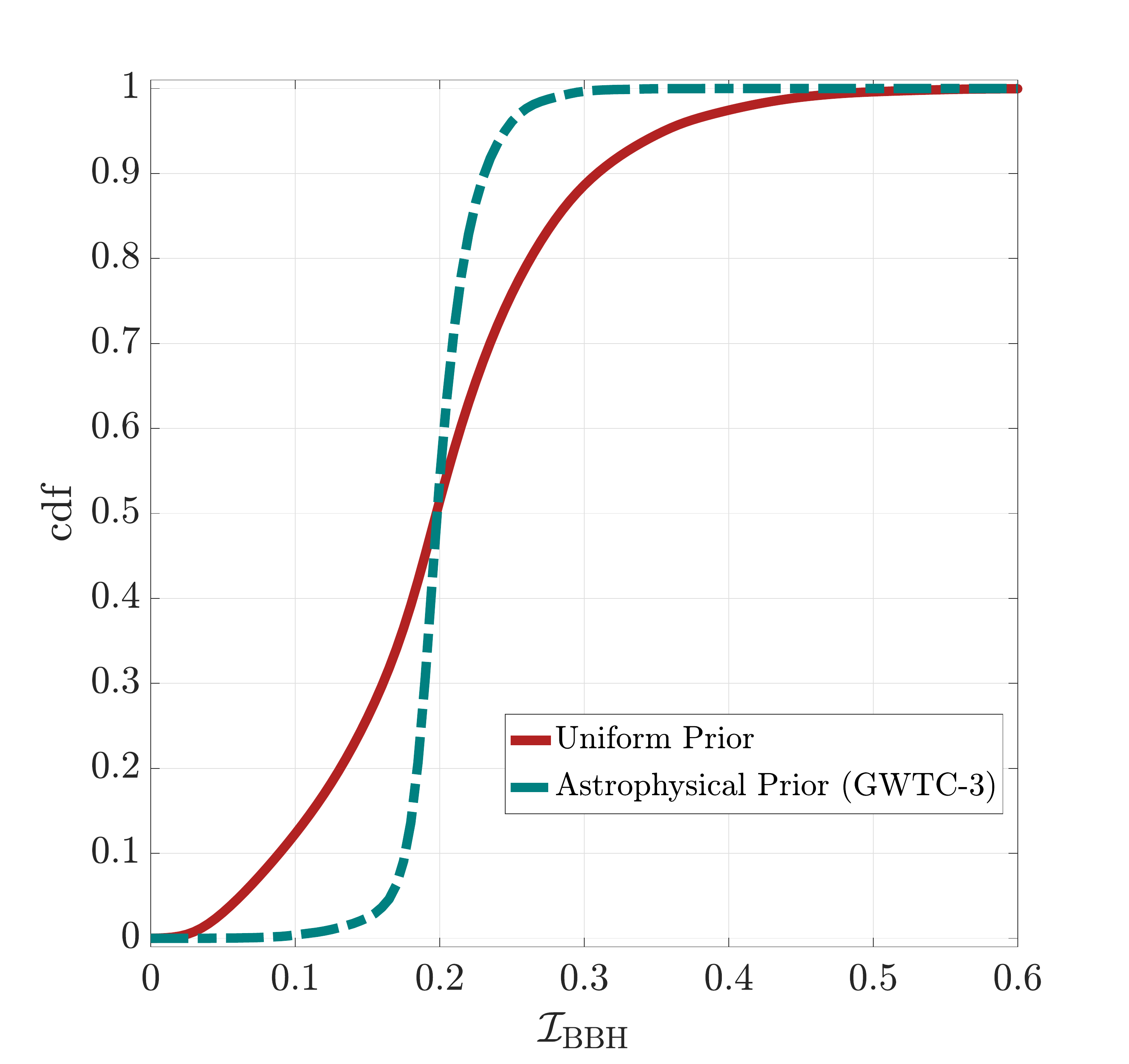}
\caption{Cumulative distribution of Merger Entropy Index for a uniform prior within the intrinsic BBH parameter space $\Lambda$ (red curve). The dotted green curve is for region of $\Lambda$ derived from the observed population of BBH by LIGO-Virgo~\cite{O3Rates}}
\label{fig2} 
\end{figure}
In Fig. \ref{fig2}, we show the cumulative distribution of $ \mathcal{I}_\mathrm{BBH}$ across the intrinsic BBH parameter space $\Lambda$. The red curve is induced by a uniform distribution within mass-ratios {$1/q\in[1,20]$}, spin magnitudes $\chi_{1,2} \in [0,0.99]$ and cosine of the spin orientations $\cos{\theta_{1,2}} \in [-1,1]$. This induced distribution is a good representation of the default prior on the Merger Entropy Index for quasi-circular BBH mergers in General Relativity.
When assuming the uniform prior, $ \mathcal{I}_\mathrm{BBH}\lesssim0.5$ at 99\% credibility, greatly restricting its allowed range. This is consistent with Fig. \ref{fig1}, which shows that higher values of $ \mathcal{I}_\mathrm{BBH}$ are confined in a small domain of $\Lambda$. 

\begin{figure}[t!]
\includegraphics[scale=0.21,trim = {0 20 0 0}]{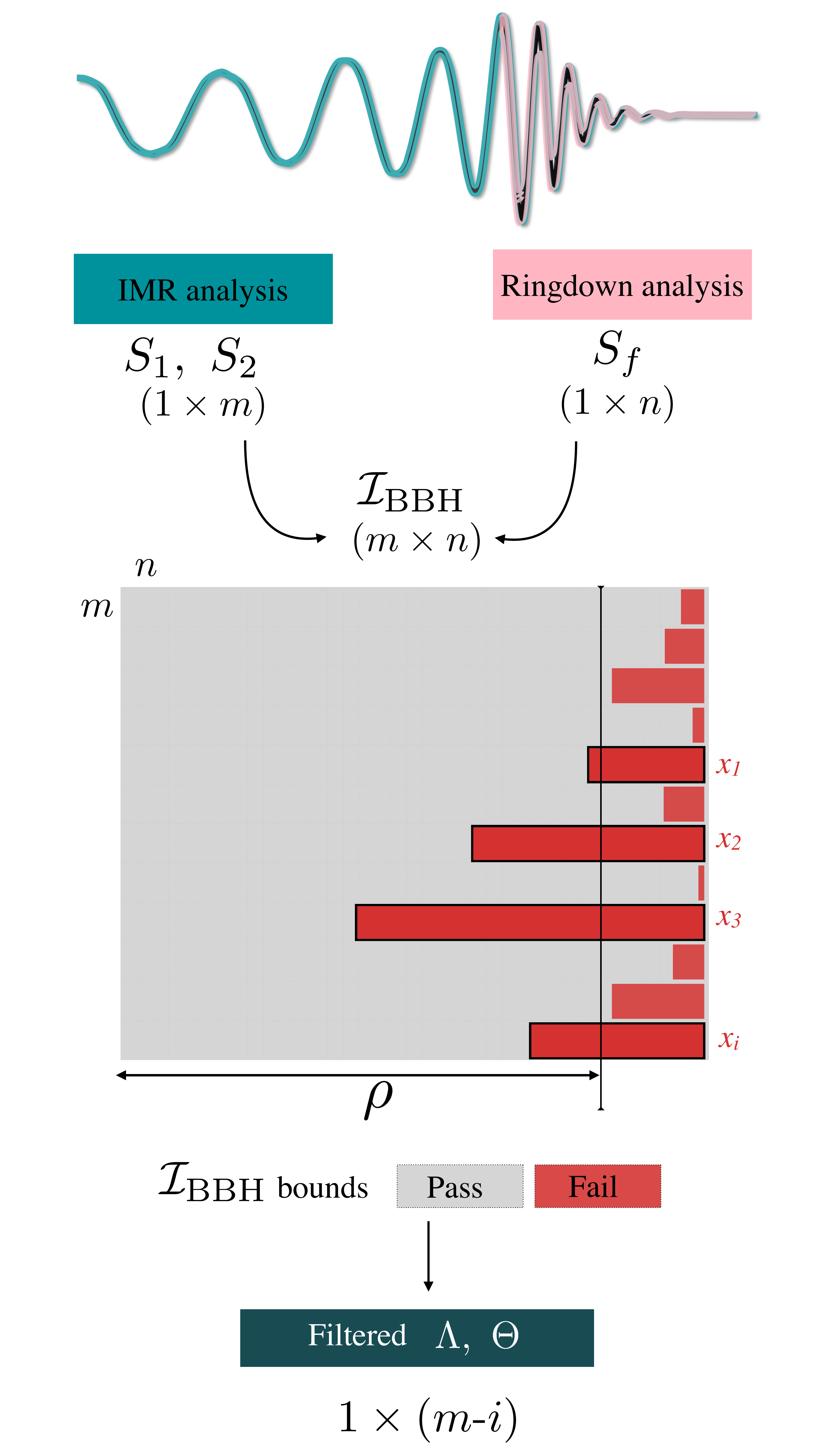}
\caption{Schematic diagram of \texttt{BRAHMA} Framework. We use a BBH detection (wavelet at the top) whose pre- and post-merger states have been independently measured. From this information we compute the Merger Entropy Index matrix. By imposing the bounds of $\mathcal{I}_\mathrm{BBH}$ stated in Equation~(\ref{bounds}, we filter parts of the inspiral posterior distribution (rows in the figure) which {\it pass} this criterion with $\geq \rho\%$.
}
\label{fig3} 
\end{figure}

The dotted green curve in Figure~\ref{fig3} shows a distribution of $ \mathcal{I}_\mathrm{BBH}$ drawn out of an {\it astrophysical prior}. This prior is derived directly from the constraints on the masses and spins for the observed population of BBH mergers in GWTC-3~\cite{O3Rates} from the LIGO-Virgo detectors. In particular, we use the distribution $p(q)\propto q^\beta$ from \texttt{PowerLaw+Peak Model} to fetch the mass-ratio, and the \texttt{Gaussian Spin Model} for the distribution of black hole spin magnitudes and orientations. For the astrophysical prior, we find $\mathcal{I}_\mathrm{BBH}$ is steeply centered around 0.2. There are no values $\gtrsim0.3$ with 99\% confidence interval. 

From the distributions discussed above, we can infer the Merger Entropy Index {has an upper-bound} as:
\begin{equation}
   0\leq \mathcal{I}_\mathrm{BBH} \lesssim  
    \begin{cases}
    1.0 ~~~ \texttt{Theoretical Limits; } \\
    0.5 ~~~ \texttt{Uniform Prior;} \\
    0.3 ~~~ \texttt{Astrophysical Prior} 
    \end{cases}
    \label{bounds}
\end{equation}

While the theoretical and uniform prior bounds of $\mathcal{I}_\mathrm{BBH}$ in Eq.~(\ref{bounds}) are independent of the BBH origins, the astrophysical bounds could well vary between the supermassive and stellar mass BBH population, and would be interesting to explore in future work. We found that KS-convergence can confidently differentiate (with a probability of one part in billion) between the LIGO-Virgo derived distribution of $\mathcal{I}_\mathrm{BBH}$~\cite{O3Rates} from the uniform prior distribution. This suggests $\mathcal{I}_\mathrm{BBH}$ can be an effective tool to test astrophysical formation channels. 

\begin{figure*}[t!]
\includegraphics[scale=0.23,trim = {0 100 0 50}]{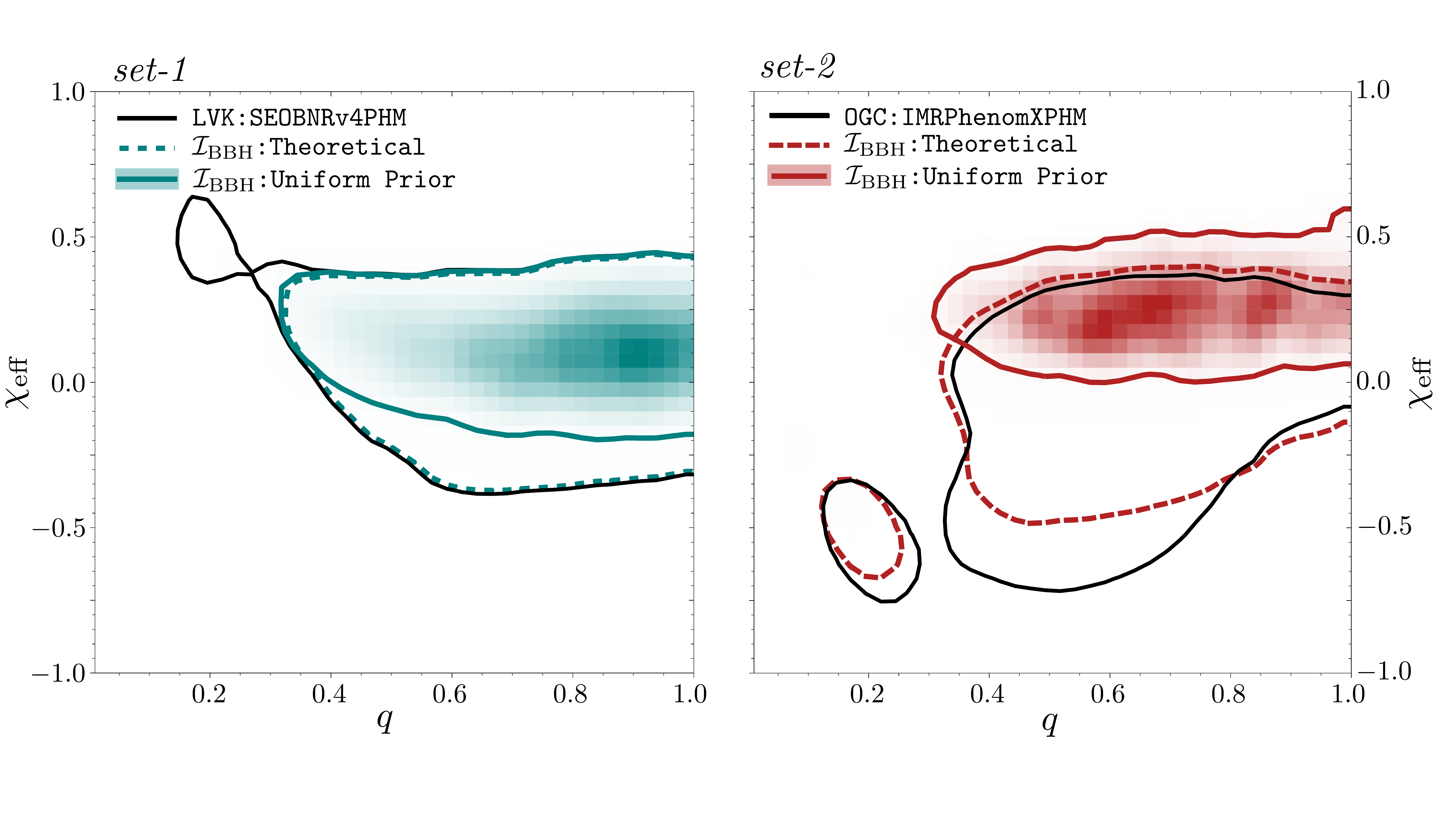}
\caption{New constraints on GW190521's mass-ratio and effective spins using \texttt{BRAHMA} {\it Left:} Constraints using the pre- and post-merger measurement from {\it set-1}, i.e. the LIGO-Virgo-KAGRA Collaboration (LVK). The black curve refers to the published distribution from the {SEOBNRv4PHM} model. The dashed green line shows the constraints we obtain in \texttt{BRAHMA}  by imposing theoretical bounds of $\mathcal{I}_\mathrm{BBH}$. The filled green contour is the constraint from \texttt{BRAHMA} for a uniform prior bound of $\mathcal{I}_\mathrm{BBH}$. {\it Right:} Constraints using pre- and post-merger measurement from {\it set-2}, i.e. the Open Gravitational-Wave Catalog (OGC). The black curve refers to the published distribution from the {IMRPhenomXPHM} model. The red dashed and red contour refers to the constraints obtained from \texttt{BRAHMA}  by imposing theoretical and uniform prior bounds of $\mathcal{I}_\mathrm{BBH}$ on this set. The two-dimensional curves show the 90\% credible regions. }
\label{fig4} 
\end{figure*}

\prlsec{Novel GW Framework} 
By adopting the Merger Entropy Index, we introduce a generalized framework called \textbf{BRAHMA} (\textbf{B}ina\textbf{R}y bl\textbf{A}ck \textbf{H}oles \textbf{M}erger entropy \textbf{A}nalysis).
The schematic flow of \texttt{BRAHMA} is shown in Fig.~\ref{fig3}. Every BBH detection whose pre- $(S_1, S_2)$ and post-merger $(S_\mathrm{f})$ entropies have been independently measured (e.g. because models with different physical content have been used to infer them) have to satisfy the upper and lower bounds of the theoretical limits of $\mathcal{I}_\mathrm{BBH}$ from Eq.~(\ref{bounds}). This is under the assumption that the observed BBH population is consistent with the predictions of General Relativity, which by now has a strong experimental support~\cite{Carullo:2021dui, 2021Isi, O3aTGR}. With the current generation of gravitational-wave data analysis tools, the pre-merger entropy will typically be measured by conducting a Bayesian analysis over the intrinsic $\Lambda$ (masses, spins) and extrinsic $\Theta$ (distance, orientation, sky-location) BBH parameter space with inspiral-merger-ringdown (IMR) waveform models (green wavelet in Fig.~\ref{fig3}). For the post-merger entropy, our framework strictly requires an independent measurement of the black hole ringdown (pink wavelet in Fig.~\ref{fig3}), which is achieved by an independent measurement of the post-merger properties~\cite{Carullo:2021dui, 2021Isi, O3aTGR}. See Ref.~\cite{Carullo:2021yxh} for thermodynamical constraints on the information emission process of LIGO-Virgo BHs using post-merger measurements.

From the posterior distributions of pre-merger entropies (with size $1\times m$) and post-merger entropies ($1\times n$), we compute a Merger Entropy Index matrix $(m\times n)$. Assuming the detected BBH is consistent with General Relativity, every element in the matrix has to satisfy (pass) theoretical bounds stated in Equation~(\ref{bounds}). For illustration, we show them as grey boxes of $\mathcal{I}_\mathrm{BBH}$ in the matrix in Figure~\ref{fig3}. The matrix may contain some {\it failed} elements, i.e they do not obey the $\mathcal{I}_\mathrm{BBH}$ bounds (shown in the Fig.~\ref{fig3} as red boxes). For every row in the $\mathcal{I}_\mathrm{BBH}$ matrix, we compute the the fraction of {\it passed} columns to that of the total elements in the column $(n)$. If this fraction is less than the threshold $\rho$, we discard this row vector and filter the pre-merger distribution to a new length $1\times(m-i)$. As a default setting in our framework, we fix $\rho=90\%$. BRAHMA directly filters the parameter space constrained by the IMR model. All the BBH inspiral parameters, including extrinsic $(\Theta)$ ones, are now further constrained than the original output from the Bayesaian analysis, even though \texttt{BRAHMA} does not explicitly require any direct information on most of these parameters (e.g. luminosity distance, coalescence phase). Naturally, the constraints on BBH parameters are tighter if the bounds on the Merger Entropy Index are set from a uniform or astrophysical prior, compared to the most general theoretical limit. 

The framework is built on the idea that for certain systems, post-merger analyses provide a more robust measurement of the end state properties than full-signal analyses. This situation can happen when full-signal models do not correctly represent complete solutions of the Einstein equations (i.e. they are "unfaithful" to accurate numerical simulation of BBHs), or when they lack the inclusion of certain physical effects. 
The post-merger stage is generally simpler to model regardless of the pre-merger history of the BBH system, and that the agnostic post-merger models employed in our analysis do not rely on numerical calibration to quasi-circular mergers.
For example, eccentricity effects are much less relevant in the post-merger due to fast binary circularisation in the late stages of the pre-merger dynamics, see Fig.~4 of~\cite{Huerta:2019oxn}. 
On the contrary, pre-merger models employed by the LVK collaboration currently cannot incorporate astrophysically interesting scenarios such as eccentric orbits, dynamical captures and high-mass ratio precessing binaries.
Some progress in some of these directions has been recently obtained using Effective One Body models for non-precessing eccentric systems and eccentric dynamical captures~\cite{Nagar:2020xsk, Gamba:2021gap}. However, such templates currently lack the inclusion of BHs spins and higher harmonics.

\prlsec{Applications to GW190521} To demonstrate the application of our \texttt{BRAHMA} framework, we apply it to the parameters measurements of GW190521. This event has been shown to suffer from systematic uncertainties of its pre-merger properties~\cite{Romero-Shaw:2020thy, Bustillo:2020ukp, Bustillo:2020syj, Gamba:2021gap}, thus falling in the category of systems to which our framework is especially relevant.
Detected during the third observing run of the LIGO and Virgo detectors, this event marks the first direct measurement of an intermediate-mass black hole~\cite{2020PhRvL.125j1102A}, though its astrophysical formation remains so far mysterious~\cite{190521_Astro}. Such massive BBHs have an advantage of an independent measurement of the remnant mass and spin $(m_\mathrm{f}, \chi_\mathrm{f})$ from the ringdown stage~\cite{2019Gregmodes}. Several studies have performed a full Bayesian inference on the pre- and post-merger properties of GW190521 using a variety of gravitational waveform models and numerical relativity simulations~\cite{190521_Astro, 2020ApJ...903L...5R, 2021Nitz, 2021arXiv210506360E, 2020arXiv200905461G}. From these data products, we chose two fully independent set of measurements to test \texttt{BRAHMA}:

\begin{enumerate}[wide, labelwidth=!, labelindent=0pt]
\item[{\it set-1:}] We choose the waveform model SEOBNRv4PHM~\cite{2020SEOB} to measure the pre-merger entropies $(S_1, S_2)$. We do so using its posteriors on detector-frame masses $(m_1,m_2)$ and spin magnitudes $(\chi_1,\chi_2)$, available from the Gravitational-Wave Open Science Center (GWOSC) maintained by the LIGO-Virgo-KAGRA Collaboration~\cite{GWOSC}. These pre-merger constraints were obtained with the Bayesian parameter scheme \texttt{RIFT}~\cite{lange2018rapid}. For the post-merger entropy $(S_\mathrm{f})$, we use the posteriors on the remnant mass and spin $(m_\mathrm{f}, \chi_\mathrm{f})$ obtained through the \texttt{pyRing} analysis framework~\cite{pyRing} with the \texttt{Kerr\_222} model~\cite{Giesler:2019uxc, 190521_Astro}. This template incorporates the $\ell=m=2$ quasi-normal modes up to the second overtone ($n=0,1,2$), and is applied starting at the peak of the waveform (a proxy for the {\it merger}). Among the post-merger models employed in~\cite{190521_Astro}, it provides the most precise measurements of ringdown, and thus that of remnant mass and spin.

\item[{\it set-2:}] From the Open Gravitational Catalog (OGC)~\cite{3OGC}, which is an independent analysis on the open data,  we get the pre-merger estimates for the IMRPhenomXPHM waveform model~\cite{IMRPhenomXPHM}. Their analysis is performed using the Bayesian code \texttt{pyCBC Inference}~\cite{PyCBCInf}. The post-merger estimates are obtained from a Kerr template~\cite{LIGOScientific:2020tif} incorporating the $(\ell,m,n) = \{(2,2,0), (3,3,0)\}$ modes, as there is claimed evidence for both these ringdown modes in GW190521~\cite{2021arXiv210505238C}.
\end{enumerate}

\begin{table}[t!]
\begin{ruledtabular}
\begin{tabular}{l c c | c c }

 & \multicolumn{2}{c}{\it set-1:}  &  \multicolumn{2}{c}{\it set-2:} \\

 & \texttt{LVK}  & \texttt{With}  & \texttt{OGC} & \texttt{With} \\
Parameter & \texttt{SEOB-PHM}  & \texttt{BRAHMA} & \texttt{IMR-XPHM} & \texttt{BRAHMA} \\
\hline
\rule{0pt}{3ex}%

Lower-bound on $q$ & $0.32$ & $0.39$ & $0.22$ & $0.37$   \\

\rule{0pt}{3ex}%

Lower-bound on $\chi_\mathrm{eff}$  & $-0.28$ & $-0.12$ & $-0.60$ & $+0.06$ \\

\rule{0pt}{3ex}%
Probability($z \leq$ AGN) &  $10\%$ & $7\%$ & $41\%$ & $5\%$  
\rule{0pt}{3ex}%
\end{tabular}
\end{ruledtabular}
\caption{Source properties for GW190521 with $\mathcal{I}_\mathrm{BBH}$ framework and imposing uniform prior. For mass-ratio $q$ and effective inspiral spin $\chi_\mathrm{eff}$, we provide the lower-bound at the 90\% confidence interval. The third row is the cumulative probability for the inferred redshift $z$ to be equal or less than the one from the claimed AGN flare, J1249+349~\cite{2020ZTF}. 
}
\label{tab:parameters}
\end{table}

\begin{figure}[t!]
\includegraphics[scale=0.23,trim = {50 40 0 0}]{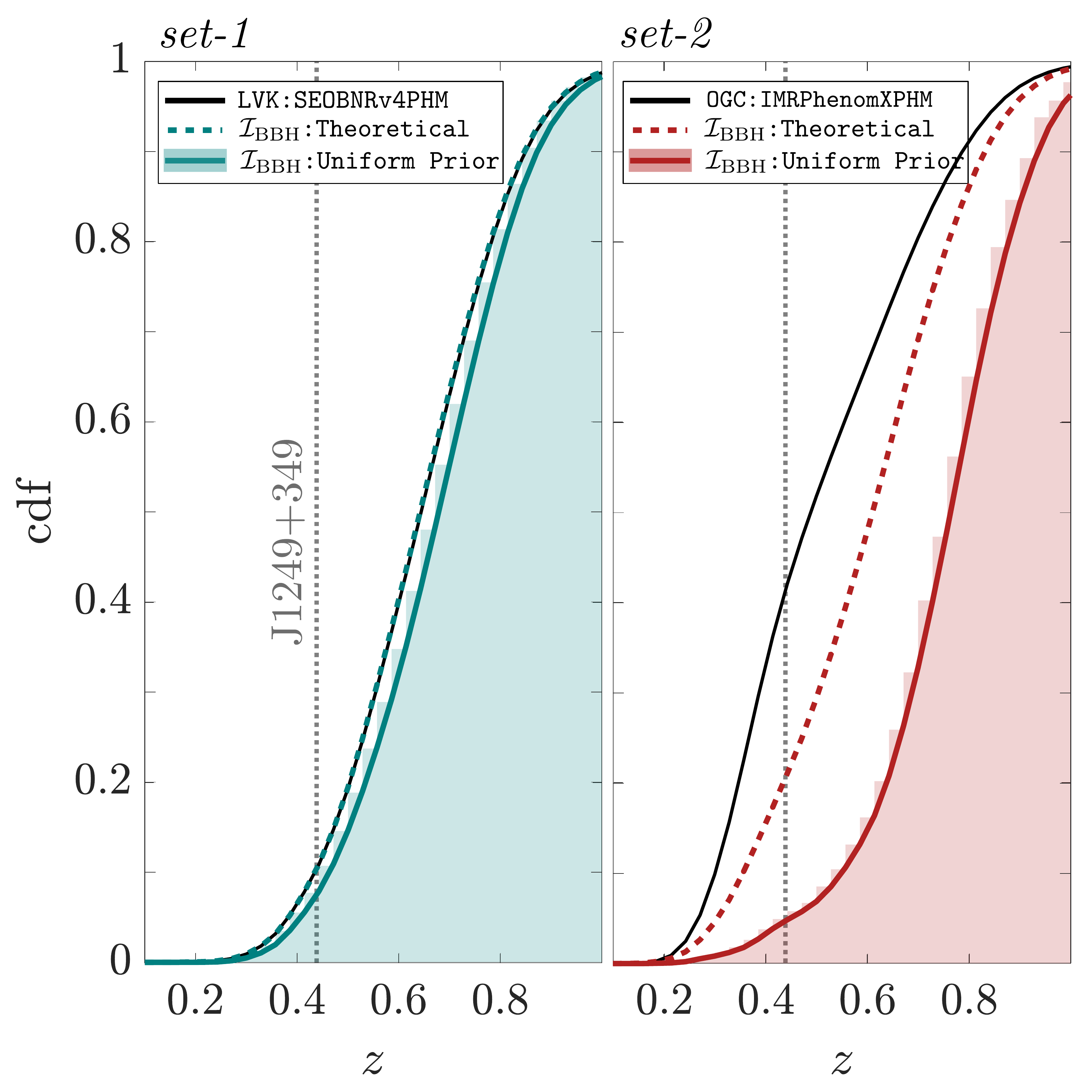}
\caption{New constraints on the redshift of GW190521 using the \texttt{BRAHMA} framework. The color coding on both the panels is same as Figure \ref{fig4}. The dotted grey line refers to the redshift of the electromagnetic candidate observed by the ZTF collaboration~\cite{2020ZTF}.  }
\label{fig5} 
\end{figure}

As shown in the two panels of Fig.~\ref{fig4}-\ref{fig5}, the inference of pre-merger properties by these two sets is significantly different (compare the black curves in left-right panels). However, both the sets provide support for (i) an asymmetric mass distribution $(q\lesssim0.3)$, (ii) negative effective inspiral spins $(\chi_\mathrm{eff}\lesssim -0.2)$ and (iii) a source redshift that overlaps with an AGN flare, J1249+349~\cite{2020ZTF}. These features carry strong implications for the astrophysical origins of GW190521. By applying our \texttt{BRAHMA} framework, we find that both {\it set-1} and {\it set-2} no longer provides evidence for the features (i)-(iii). Table \ref{tab:parameters} summarizes the overall improvement in pre-merger properties of GW190521 using \texttt{BRAHMA}. 

After applying the \texttt{theoretical} prior of $\mathcal{I}_\mathrm{BBH}$ in \texttt{BRAHMA}, {\it set-1} (\texttt{LVK:SEOB-PHM}) no longer shows the bimodality in $q-\chi_\mathrm{eff}$ parameter space. For {\it set-2} (\texttt{OGC:IMR-XPHM}), the bimodality still remains, however, the support for $\chi_\mathrm{eff}\lesssim0$ is significantly lower (see dashed lines in Fig. \ref{fig4}). These bounds also have significant impact on the source redshift for {\it set-2}, dropping its cumulative probability at $z\leq0.438$ (electromagnetic candidate claim) by a factor of ${\sim}2$, confirming GW190521 is farther than the claimed electromagnetic candidate. 

When the \texttt{uniform prior} bounds are adopted in \texttt{BRAHMA}, we see a significant reduction in the 90\% confidence intervals  in both the sets.
The most notable changes appear in our constraint from {\it set-2} on the effective inspiral spin of GW190521, which suggests the parameter is positive with 90\% confidence. Also, for the same set the cumulative probability for source redshift at $z\leq0.438$ drops to ${\sim}5\%$ (a factor of 8). Filtered posteriors from both sets no longer show any bimodality, and in fact, converge to similar a post-merger distribution after applying our $\mathcal{I}_\mathrm{BBH}$ framework. Moreover, the distribution from these two sets now seems consistent with other waveform models that have analyzed GW190521 \cite{190521_Astro}, suggesting a possible improvement in waveform systematic with $\mathcal{I}_\mathrm{BBH}$. 

By imposing the \texttt{astrophysical prior} in \texttt{BRAHMA}, we find that 99.9\% (95\%) of pre-merger distribution is filtered out from {\it set-2} ({\it set-1}). This provides an additional confirmation that GW190521 is at the tail end of the astrophysical population derived from all other LIGO-Virgo black holes~\cite{O3Rates}.

\prlsec{Conclusion}
We introduce a new thermodynamical formalism into gravitational-wave astrophysics. 
For the current generation detectors, we find our formalism is particularly powerful for probing GW190521-like lower-range intermediate-mass black holes $({\sim}10^{2-3})~M_\odot$ ~\cite{2020NatAs...4..260J, O3IMBH}. Our framework is applicable to BBH lying in regions of the parameter space where full numerical simulations are not extensively available and for which the pre- and post-merger states can be independently measured, for all mass ranges. Our formalism will be even more germane for next-generation detectors Cosmic Explorer~\cite{2019arXiv190704833R}, Einstein Telescope~\cite{Punturo:2010zz}, upcoming space-mission LISA~\cite{2017arXiv170200786A} and potential deci-Hertz experiments \cite{Sato_2017, Jani_2021}, as we expect the post-merger measurements to be more often and louder. Future work will explore the implications of the Merger Entropy Index $(\mathcal{I}_\mathrm{BBH})$ and the \texttt{BRAHMA} framework on BBH demographics, formation channels and gravitational waveform systematics.

\section{acknowledgments}
We thank Maximiliano Isi and Nathan Johnson-McDaniel for useful comments. P.H.'s research was supported by the Immersion Initiative at Vanderbilt University. K.J and K.H.B.'s research was supported by the Stevenson Chair endowment funds at Vanderbilt University.

This material is based upon work supported by NSF's LIGO Laboratory which is a major facility fully funded by the National Science Foundation. This research has made use of data, software and/or web tools obtained from the Gravitational Wave Open Science Center (\url{https://www.gw-openscience.org/}), a service of LIGO Laboratory, the LIGO Scientific Collaboration and the Virgo Collaboration. LIGO Laboratory and Advanced LIGO are funded by the United States National Science Foundation (NSF) as well as the Science and Technology Facilities Council (STFC) of the United Kingdom, the Max-Planck-Society (MPS), and the State of Niedersachsen/Germany for support of the construction of Advanced LIGO and construction and operation of the GEO600 detector. Additional support for Advanced LIGO was provided by the Australian Research Council. Virgo is funded, through the European Gravitational Observatory (EGO), by the French Centre National de Recherche Scientifique (CNRS), the Italian Istituto Nazionale di Fisica Nucleare (INFN) and the Dutch Nikhef, with contributions by institutions from Belgium, Germany, Greece, Hungary, Ireland, Japan, Monaco, Poland, Portugal, Spain.

\bibliography{references.bib}

\end{document}